\documentclass[useAMS,usenatbib]{mn2e}

\input psfig.sty

\title[QbCII: Baryonic halo sizes]{Galaxy Clusters in the Line of Sight to Background Quasars: 
II. Environmental effects on the sizes of baryonic halo sizes}
\author[Padilla \& the QbC Team]{
\parbox[t]{\textwidth}{
Nelson Padilla$^{1}$\thanks{E-mail:
npadilla@astro.puc.cl}, 
Ivan Lacerna$^{1}$,
Sebasti\'an Lopez$^{2}$,
L. Felipe Barrientos$^{1}$,\newline
Paulina Lira$^{2}$,
Heather Andrews$^{1}$,
Nicol\'as Tejos$^{2}$.
}
\\
$^{1}$Departamento de Astronom\'\i a y Astrof\'\i sica, Pontificia Universidad Cat\'olica de Chile, Santiago, Chile.\\
$^{2}$Departamento de Astronom\'\i a, Universidad de Chile, Casilla 36-D, Santiago, Chile.\\
}
\def\mgii{Mg\,{\sc ii}}
\def\mgmath{Mg\,{\sc II}}

\begin{document}

\date{Accepted 2009 February 10. Received 2009 February 09; in original form 2008 November 03}

\pagerange{\pageref{firstpage}--\pageref{lastpage}} \pubyear{2009}

\maketitle

\label{firstpage}

\begin{abstract}
Based on recent results on the frequency of \mgii\ 
absorption line systems in the ``QSO behind
RCS clusters" survey (QbC), we analyse the effects of the cluster environment on the sizes of baryonic
haloes around galaxies.  We use two independent models, i) an empirical halo occupation model which
fits current measurements of the clustering and luminosity function of galaxies at low and high
redshifts, and ii) the GALFORM semi-analytic model of galaxy formation, which follows the evolution
of the galaxy population from first principles, adjusted to match the statistics of low and
high redshift galaxies.  
In both models we constrain the \mgii\ halo sizes of field and cluster galaxies
using observational results on the observed \mgii\ statistics.
Our results
for the field are in good agreement with previous works, indicating a typical \mgii\ halo size of 
$r_{\mgmath}\simeq 50$h$_{71}^{-1}$kpc in the semi-analytic model, and slightly lower in the halo occupation
number approach.  For the cluster environment, we find that both models require a median \mgii\ halo size
of $r_{\mgmath}< 10$h$_{71}^{-1}$kpc in order to reproduce the observed statistics on absorption line
systems in clusters of galaxies.
Based on the Chen \& Tinker (2008) result that stronger systems occur closer to the 
\mgii\ halo centre,
we find that 
strong absorption systems in clusters of galaxies 
occur at roughly a fixed fraction of the cold-warm halo size out
to $1$h$_{71}^{-1}$Mpc from the cluster centres.  In contrast, weaker
absorption systems appear to occur at progressively shorter relative fractions of this halo
as the distance to the cluster centre decreases.  These results 
reinforce our conclusions from Paper I and
provide additional independent support for the stripping 
scenario of the cold gas of galaxies in massive clusters
by the hot intracluster gas, e.g., as seen from X-ray data.
\end{abstract}

\begin{keywords}
galaxies: clusters, galaxies: general, galaxies: structure, galaxies: haloes
\end{keywords}

\section{Introduction}

Understanding the influence of environment on galaxy evolution either in
the field or in galaxy clusters, provides important missing clues on the
interplay of internal and external mechanisms that shape the galaxy
population. In this work we concentrate on the statistics of
\mgii\ absorption systems in galaxies associated with high-redshift
clusters, as recently obtained by Lopez et al. (2008; hereafter 'Paper
I'). Our aim is to confront those statistics with different galaxy models in
order to constrain the sizes of the cold-warm baryonic component of 
cluster galaxies\footnote{ The temperature of the \mgii\ gas is expected
to be a few $10^4$ K, and therefore is termed here cold-warm
gas.  Notice that in some other cases, e.g., in Chen \& Tinker (2008), \mgii\ is
referred to as cold gas.}.

Ever since the first studies of QSO absorption lines, \mgii\  was recognized as an
excellent tracer of high-redshift galaxies (Bergeron \& Stasinska 1986;
Petitjean \& Bergeron 1990; Steidel \& Sargent 2002; Churchill et al. 2000;
Lanzetta, Turnshek \& Wolfe, 1987, Tytler et al., 1987, Steidel \& Sargent,
1992). These seminal works were then followed by numerous studies 
(Churchill et al. 1999; Nestor, Turnshek \& Rao. 2005; Nestor, Turnshek \&
Rao 2006, Narayanan et al. 2007, Lynch, Charlton \& Kim 2006; Prochter,
Prochaska, \& Burles 2006) that were able to distinguish between weak and
strong \mgii\ absorption systems as possibly two distinct populations. 
When selecting the stronger \mgii\ systems (with rest-frame 
equivalent width $W^{2796}_0>0.3$ \AA) the associated
galaxies present a slight preference for blue, starburst galaxies with
high metallicities (Zibetti et al., 2007, Ellison, Kewley \&
Mall\'en-Ornelas, 2005). However, it is still not clear whether the 
stronger  systems really occur in extended (hundreds of kpc) and virialised
halos already in place at $z=2$ (Churchill et al. 2005; Steidel et al. 2002)
or if some are the signature of violent galaxy-scale outflows (Bouche et
al. 2006; Prochter et al. 2006).

Despite these extensive absorber-galaxy studies, and perhaps due to our
lack of knowledge about the nature of \mgii\ absorbers, it has been difficult so
far to establish correlations between absorption-line strength and
absober-galaxy luminosity and impact parameter.  Although some tentative
correlations were found in the past (Churchill et al., 1999; Steidel et
al. 1995), it is now becoming clear from deep imaging of the QSO fields, that
not only the \mgii\ covering fraction must be less than unity (Chen \&
Tinker, 2008, Kacprzak et al., 2008), contrary to what was assumed for
years, but also that there must be correlations with the galaxy
environment.

In Paper I we proposed to study the environmental
dependence of the size of the cold-warm baryonic component using the
frequency of \mgii\ in the line-of-sight (LOS) to QSOs close in projection
to foreground clusters. Thus, contrary to previous studies, our samples were constructed
in an unbiased way by
selecting a set of known galaxy overdensities and then searching
for absorption systems in their neighbourhoods.  In
these particular lines of sight 
we obtained a comparatively flatter equivalent width distribution 
with respect to what has been
obtained for the field. This result suggests that the cluster environment
could be pruning the baryonic halo size significantly.

Generally, the interpretation of the statistics on \mgii\ absorption systems
in the field\footnote{ In this paper we refer to ``the field" as a
collection of objects which has been drawn at random from the full galaxy
population, and therefore includes many different environments, from very
underdense regions to high mass clusters.  This coincides with the
selection of QSOs made by several authors to detect \mgii\ absorption
systems, and will be adopted in our analysis of the numerical models.} has
been done following analytic approaches (as in Churchill et al., 1999).
Only more recent works (Tinker \& Chen, 2008, Chen \& Tinker, 2008) use a
model motivated by observational data within a cosmological framework, the
Halo Occupation Model (HO, Seljak, 2000, Scoccimarro et al. 2001; Berlind \&
Weinberg 2002; see Zheng \& Weinberg 2007, and  references therein), to study
the origin of the \mgii\ absorption systems, their geometry, and the
dependence of equivalent width with the distance from a \mgii\ halo centre
\footnote{The HO models which we refer to in this paper,  are rather
complex Monte-Carlo simulations that reproduce density profiles of haloes
of different sizes via fits to numerical simulation results, and populate
these haloes with baryonic components following observational measurements
of statistics such as the luminosity and correlation functions.  All other
non-HO models assuming statistical distributions for observational
quantities will be referred to simply as Monte-Carlo simulations.}.

Building upon the results presented in Paper I, we perform a
theoretical study of \mgii\ absorption systems using models to determine how
the size of the cold-warm baryonic halo depends on environment.  In order to
do this we use two different models, (i) a HO model, constructed
empirically  by Cooray (2006), adopting a $\Lambda$CDM cosmology, and
(ii) a semi-analytic model by Bower et al. (2006, a version of the GALFORM
model by Cole et al., 2000) which follows the evolution of galaxies in a
$\Lambda$CDM Universe from first principles, tuned to reproduce the observed
galaxy population at a wide range of redshifts (other semi-analytic models
include Baugh et al., 2005, Croton et al., 2006, Lagos et al., 2008, among
others). We then concentrate on determining the cold-warm baryonic
halo sizes and their dependence on  galaxy environment, without modeling the
distribution of gas in a direct way, but instead establishing the required
cold-warm halo size needed to reproduce the observed \mgii\ absorber 
statistics presented in Paper I for clusters of galaxies.  For simplicity,
we start assuming a unity covering fraction and then discuss the
implications of this choice.

Throughout this paper the cosmological model adopted
is characterised by the ``concordance" parameters in line with estimates from
the large-scale distribution of galaxies and the temperature fluctuations in the cosmic
microwave background (S\'anchez et al., 2006), namely,
$\Omega_m=0.25$, $\Omega_\Lambda=0.75$,  $\Omega_{b}=0.045$,
a Hubble constant $H_0=71\times$h$_{71}$km$/$s$/$Mpc, with h$_{71}=1$, $n=1$ and $\sigma_{8}=0.9$.

The paper is organised as follows.  In Section 2 we briefly review the
observational results  from Paper I and the additional refinements applied
to them to take into account the contamination from the field and the
large-scale structure. Section 3 presents the two models in detail.
Section 4 
describes the results, and a discussion is presented in Section 5.  Our
conclusions are summarised  in Section 6. 

\section{Observations}

In this section we describe the available datasets used for the present
work and the corrections we have made to account for contamination from the field and large-scale 
structures.

\subsection{The data}

In Paper I we used the Red sequence Cluster Survey (RCS, Gladders \& Yee, 2005) 
to perform a search of clusters 
in the LOS to background quasars, and constructed two different sets
of cluster-QSO pairs.  
The first set includes $46$ high-resolution QSO spectra, and the second comprises
lower resolution spectra (these numbers do not include restrictions on signal-to-noise
ratios).  
In all cases, the maximum physical distance between
the line-of-sight to QSOs and the cluster centres was set to $2$h$_{71}^{-1}$Mpc, for
a total sample of $529$ cluster-QSO pairs.

We estimate the median mass
of RCS clusters in the sample using the B$_{\rm gcR}$ parameter (Gilbank et al., 2007) 
which can be used to estimate
individual cluster masses. 
For this work we only take into account results from clusters 
with at least a $2-\sigma$ detection of B$_{\rm gcR}$ (that is, $2-\sigma$ away from zero).  
For this subsample of RCS clusters (all of which are paired to a background QSO for a total of
212 pairs) we
find a median mass of $(1.64^{+0.856}_{-0.345})\times 10^{14}$h$_{71}^{-1}M_{\odot}$ 
(errors correspond to 
the $20$ and $80$ percentiles of the mass distribution) which we will use in the remainder of this work.

Clusters in the RCS survey are identified
using a likelihood method which fits simultaneously for a projected cluster
density profile and a red sequence at different redshifts (see Paper I for more details);
this method produces an estimate of the cluster photometric redshift as a by-product.
A \mgii\ system is associated to a RCS cluster when
the redshift of the absorption falls within
a $1-\sigma$ interval around the cluster photometric redshift.  In some cases
a given absorption system can satisfy this criterion for more than one cluster,
in which case we test two different options, to assign the absorption system
(i) to the closest cluster in projection (referred to as "nearest cluster", 
and (ii) to the farthest cluster in projection ("farthest cluster").
We will show that this choice does not affect our results in a significant way.
respectively.  

Our subsamples are defined via
lower limits of equivalent width (EW).  This is particularly
important since without a clear modeling of the dependence of the line strength in our
models we cannot exclude the occurrence of strong lines, particularly when
taking into account indications that these occur closer to the centre of the absorbing galaxy
(Chen \& Tinker, 2008).
We construct (i) sample S:WS
(weak and strong systems) which includes
all QSO-Cluster pairs from the high resolution spectra,
and (ii) sample S:St (strong systems only) which consists of QSO-Cluster
pairs selected from the low-resolution data.  
Notice that sample S:St includes the QSO-Cluster pairs from S:WS (all the
available pairs), but only considers absorption systems with EW above the lower limit of $1$\AA.
We define hits in the same fashion as in Paper I, i.e., absorption systems within 
$\pm \Delta z$ of a cluster redshift, where $\Delta z$ is the uncertainty in the
cluster redshift. 

Notice that in order to maximize the number of hits, we do not impose the constrains on 
EW completeness or on pair redshifts used in paper I to correct for cluster completeness. 
As we show below, these more relaxed sample definitions introduce variations in our
results that are negligible when compared to the field and clustering corrections.

    \begin{table*}
    \begin{center}
    \begin{tabular}{ccccc}
    \hline
    \hline
    \noalign{\vglue 0.2em}
    range in $b/$h$_{71}^{-1}$Mpc  & $N_{hits}$ & Field corrected & Clustering corr. & Number of QSO-Cluster pairs \\
    \noalign{\vglue 0.2em}
    \hline
    \noalign{\vglue 0.2em}
     $0.0$ to $0.5$ & $8$  &$4.79$&$4.22$&11\\
     $0.5$ to $1.0$ & $5$  &$2.41$&$2.05$&12\\
     $1.0$ to $1.5$ & $1$  &$0.47$&$0.37$&13\\
     $1.5$ to $2.0$ & $1$  &$0.48$&$0.40$&10\\
    \noalign{\vglue 0.2em}
    \hline
    \hline
    \end{tabular}
    \vskip -0.2cm
    \caption{Number of \mgii\ doublet hits in the S:WS sample (second
column) for different ranges of impact parameter to the cluster centre (indicated in the first column).
The third column contains the number of hits corrected from the contribution from the
Field; the fourth column includes the effects of clustering for the correction; the fifth column
shows the total number of QSO-Cluster pairs.}
    \end{center}
    \end{table*}\label{table:hits}

    \begin{table*}
    \begin{center}
    \begin{tabular}{ccccc}
    \hline
    \hline
    \noalign{\vglue 0.2em}
    range in $b/$h$_{71}^{-1}$Mpc  & $N_{hits}$ & Field corrected & Clustering corr. & Number of QSO-Cluster pairs \\
    \noalign{\vglue 0.2em}
    \hline
    \noalign{\vglue 0.2em}
     $0.0$ to $0.5$ & $1$  &$0.96$&$0.95$&14\\
     $0.5$ to $1.0$ & $7$ &$6.73$&$6.70$&32\\
     $1.0$ to $1.5$ & $2$  &$1.92$&$1.91$&55\\
     $1.5$ to $2.0$ & $1$  &$0.95$&$0.94$&66\\
    \noalign{\vglue 0.2em}
    \hline
    \hline
    \end{tabular}
    \vskip -0.2cm
    \caption{Number of \mgii\ doublet hits in the S:St sample.  Columns are as in Table 1.}
    \end{center}
    \end{table*}\label{table:hits2}

\subsection{Correcting for systematic effects in the Cluster-QSO pair information}

The number of hits, $N_{hits}$,
varies with the impact parameter to the cluster centre (see Tables 1 and 2).  The number of QSO-cluster pairs
is larger than the number of QSOs with measured spectra since in many cases there are two or more clusters
in the QSO line-of-sight.
Given that a \mgii\ absorption is associated to a cluster only if the redshift difference is within the cluster
photometric redshift error, $\Delta z$, there is a chance that some of these hits will have actually
taken place in the field (which corresponds to the average environment
including voids and clusters, statistically) rather than in the associated RCS cluster.  Therefore,
we make a first correction to the number of hits by setting,
\begin{equation}
N_{hits}^{fc}=N_{hits}-N_{field},
\end{equation}
where the correction from the field takes the form,
\begin{equation}
N_{field}=\frac{dN}{dz}(z)\times 2 \Delta z  ,
\end{equation}
with $z$ being the photometric redshift of the cluster, 
$2 \Delta z$ is summed over all clusters considered in a given sample, 
and $dN/dz$ is the field estimate from the literature. For S:WS, dN/dz is given by
\begin{equation}
\frac{dN}{dz} = 1.9306 + N_\ast (1+z)^\alpha \exp\left(-\frac{W_0}{W_\ast} (1+z)^{-\beta}\right)   ,
\label{Nestor}
\end{equation}
where the first term corresponds to the Churchill et al. (1999) estimate 
extended to $0.015$\AA$<W_0^{2796}<0.3$\AA, and the second term corresponds to 
the Nestor, Turnshek \& Rao (2005) estimate for
absorbers with EWs higher than $W_0$, which in this case is set to
$W_0=0.3$\AA;
their published parameter values are
$N_\ast=1.001(\pm0.132)$,
$\alpha=0.226(\pm0.170)$,
$W_\ast=0.443(\pm0.032)$,
$\beta=0.634(\pm0.097)$
(uncertainties are shown only to illustrate the accuracy of the
fit). For sample S:St $dN/dz$ is given by the second term with $W_0=1$\AA.
The adopted redshift corresponds to the median redshift of the RCS cluster sample,
median$(z)=0.6$.  The field corrections are listed in column 3 of Tables 1 and 2.

It is well known that clusters occupy biased density peaks in the distribution
of matter which are characterised by a high amplitude cluster-mass cross-correlation
function (see for instance, Croft et al., 1997, Padilla et al., 2001).  Hence, a correction has to be introduced
to take into account the enhanced matter density around clusters  which would increase the occurrence of
\mgii\ absorbers from the field in the surrounding few Mpc around each cluster.

We then proceed to calculate the
effect of clustering in the region surrounding the clusters, for which we propose a
correction of the form,
\begin{equation}
N_{hits}=N_{hits}^{fc}-N_{\xi corr},
\end{equation}
where $N_{\xi corr}$ includes the expected excess of field hits due to the
overdensity around our cluster sample,
\begin{equation}
N_{\xi corr}=I\times N_{field}  
\end{equation}
where the factor $I$ is calculated using the cluster-mass cross-correlation function as
\begin{equation}
\small
I \hskip-.05cm=\hskip-.05cm \frac{\displaystyle \int_{r_{cm}(z-\Delta z)}^{r_{cm}(z)-Y_c} \hskip -.1cm{\frac{n(b,y')}{\left<n\right>} dy} \hskip -.1cm+ \hskip -.2cm\int_{r_{cm}(z)+Y_c}^{r_{cm}(z+\Delta z)} \hskip -.1cm{\frac{n(b,y')}{\left<n\right>} dy}}
{\displaystyle \int_{r_{cm}(z-\Delta z)}^{r_{cm}(z)-Y_c} {dy} + \int_{r_{cm}(z)+Y_c}^{r_{cm}(z+\Delta z)} {dy} }-1,
\end{equation}
where  the comoving distance in the redshift path is 
$Y_c = \sqrt{r^2_{90} - b^2}$, $r_{cm}$ indicates comoving distance,
$r_{90}=1.92$h$_{71}^{-1}$Mpc is the median of the radius containing $90$ percent of the cluster galaxies for
our sample of RCS clusters,
$\left<n\right>$ is the average galaxy density, and $n(b,y')$ is the galaxy density calculated using
\begin{equation}
n(b,y')=\left<n\right> \left[1 + \xi_M(b,y')  \right].
\end{equation}
In the last expression, $\xi_M(b,y')$ is the cluster-mass cross-correlation function,
\begin{equation}
\xi_M(x,y') = \xi_M(\sqrt{x^2 + y'^2}) = \xi_M(r) = \beta \left(\frac{r}{r_0} \right)^{\gamma},
\end{equation}
where $r_0=5$h$^{-1}$Mpc and $\gamma=-1.8$ parametrise the correlation function of the mass (see for instance
Padilla et al., 2004), and 
$\beta=1.6$ represents the bias factor of haloes with the same median mass as our sample of
RCS clusters (notice the change of notation for the bias
with respect to previous works), obtained using the ``concordance cosmology" and
the Sheth, Mo \& Tormen (2001) formalism.
As we use the cross-correlation function between clusters and the mass, there
is only one factor of the bias parameter instead of the square of the bias as it would be
the case in the cluster auto-correlation function.

The effect of these two corrections can be appreciated in the number of absorption 
systems associated to clusters, for different
ranges of impact parameter with respect to the cluster centre shown in Tables 1 and 2.
Notice that the effect of the corrections is very small for S:St, reinforcing the results
from Paper I where the frequency of strong absorption systems in clusters of galaxies
was found to be significantly larger than in the field.

\begin{figure*}
\begin{picture}(330,340)
\put(0,0){\psfig{file=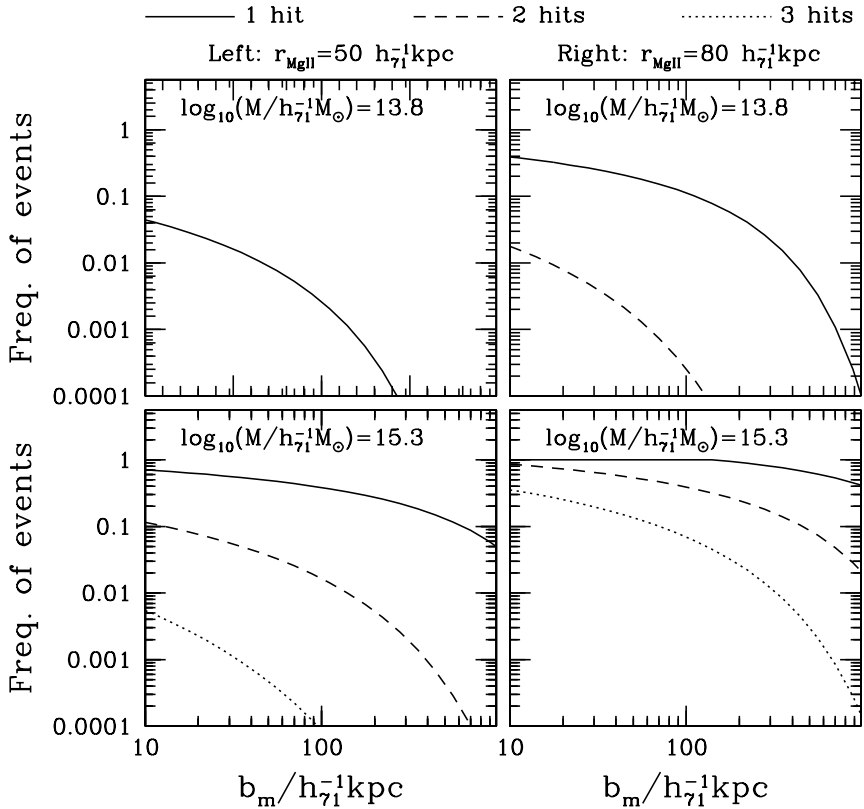,width=12.cm}}
\end{picture}
\caption{HO model:
Number of events relative to the number of QSO-cluster pairs of single, double and triple hits (solid, dashed and
dotted lines, respectively) as a function of
median impact parameter measured with respect to the cluster centre. Left
and right panels show the results from considering $r_{gal}=50$ and
$80$h$_{71}^{-1}$kpc, respectively.  The cluster mass increases from the top to the bottom
panels as indicated in the key.  
}
\label{fig:montecarlo}
\end{figure*}

\section{Models}

We use theoretical predictions from numerical simulations to estimate
the median size of the cold-warm gas halo from the available statistics on \mgii\ absorption systems from
cluster galaxies in the line-of-sight of QSOs.

In order to do this we use two different methods.  One consisting on a
Monte-Carlo simulation and another on the analysis of the output of
a semi-analytic model by Bower et al. (2006).   
The reason behind this choice is to allow our results to include a variance from
the intrinsic assumptions present in very different modeling procedures, to allow
a quantification of possible model-dependent effects.
As mentioned above, both models adopt the ``concordance" cosmology.
In the following subsections we first
present general definitions and notation, and then show the details of each of
the models.

\subsection{Notation and definitions}

In the models we will make reference to two distinct types of objects, (i) cluster-mass dark-matter haloes, 
(ii) and galaxies residing within these clusters.
In all cases, the impact parameter will refer to the projected
distance 
in physical
units at the comoving distance to the cluster
between the QSO line-of-sight and the centre of the cluster (real or simulated);
we will use the notation $b$ for this impact
parameter.  In the case of individual galaxies, we will refer to two separate quantities,
(i) the scale-length of the gaseous disk, $r_{disk}$, and (ii) the extent of the \mgii\ halo capable of producing
a \mgii\ absorption feature above a minimum equivalent width, $W^{2796}_0$, that will be accordingly specified.  
The \mgii\ halo in the model galaxies is assumed to be a uniform cloud of constant density with no holes (filling factor of unity;  see Section 5.2 for a discussion of the effects of a different value).
Finally, in all cases the observational sample to be compared with the models
will be characterised by its median equivalent width.

\subsection{Halo Occupation Monte-Carlo simulations}

The HO model is used in the literature 
to test whether the properties of individual galaxies depend on their host halo mass (e.g. 
Cooray, 2005); its parameters can be adjusted
using observed galaxy properties such as their clustering (see for instance Zehavi et al., 2004)
and luminosity functions in different bands (Cooray, 2006).  A HO model
does not include physical prescriptions for galaxy evolution, although it can be used to
infer the evolution of the population of galaxies in haloes via comparison with observations
at various redshifts.

In this first approach we use a
HO model with parameters adjusted by Cooray (2006) to match the luminosity
function of Sloan Digital Sky Survey (Abazajian et al., 2005) galaxies
brighter than r-band absolute magnitudes $M_r=-17$.
This HO model assumes that the number of galaxies per halo depends on the halo mass
such that haloes with masses above a minimum value $M_{min}=2.8\times 
10^{10}$h$_{71}^{-1}M_{\odot}$  host one central galaxy,
and the number of satellite galaxies per halo increases
according to a power law $N_{sat}=(M/M_{sat})^{\beta}$, where $M$ is the dark-matter mass
of the host halo, 
$M_{sat}=1.69^{+3.2}_{-1.5}\times 10^{13}$h$_{71}^{-1}M_{\odot}$, 
and $\beta=0.76\pm0.18$.

We combine this recipe with a
projected Navarro, Frenk \& White (1997, NFW) 
profile (from Yang et al. 2003) which we assume is followed by
galaxies populating dark-matter haloes.
The NFW profile
allows us to determine the projected density of galaxies as a function
of the distance to the cluster centre, which is scaled according
to the expected number of galaxies for clusters of a given mass, obtained
from the HO model.  The projection of the NFW profile is done by integrating
the 3-dimensional profile on the direction of the line-of-sight out to $2$
virial radii from the cluster centre (see Yang et al., 2003, for more details).
We use this projected galaxy number density, $\sigma(b)$, as a function of
the projected distance to the cluster centre to determine the
median \mgii\ halo size, $r_{\mgmath}$, necessary to produce a given frequency of hits, $N_{hits}/N_{LOS}$, in a sample
of $N_{LOS}$ QSO-cluster pairs using,
\begin{equation}
r_{\mgmath}=\sqrt{\frac{N_{hits}}{\pi \sigma(b) N_{LOS}}},
\label{rmgii}
\end{equation}
where we have assumed that all galaxies have equal \mgii\ halo sizes.

Figure \ref{fig:montecarlo} shows the expected number of hits per QSO-cluster pairs
for clusters with median masses 
of $10^{13.8}$h$_{71}^{-1}M_{\odot}$ (top panels), increasing to
$10^{15.3}$h$_{71}^{-1}M_{\odot}$ (bottom panels).  We show the results for $1$, $2$ and $3$ simultaneous
hits (solid, dashed and dotted lines) in the LOS of one target quasar as a
function of the median impact parameter, $b_m$, calculated at the median
redshift of the RCS cluster sample, median$(z)=0.6$.  
In this case, we assume two values for the median \mgii\ halo sizes, $r_{\mgmath}=50$ and
$80$h$_{71}^{-1}$kpc (left and right panels, respectively), which correspond to
a range of sizes consistent with estimates from 
absorption systems of field galaxies in QSO spectra (Churchill, 2001,
Churchill \& Vogt, 2001, Nestor, Turnshek \& Rao, 2005). The models show that it is easier to obtain more
events of multiple hits when observing more massive clusters as well as
when choosing background QSOs closer to the cluster centres.  
Also, a larger \mgii\ halo will produce more hits, as expected.
These plots allow us to study the particular case of LOS number 14 from
Paper I, where two \mgii\ absorption systems overlap with the allowed ranges of photometric
redshifts of $6$ individual RCS clusters.  
Using the expected frequency for single and double hits for the median mass of our
RCS cluster sample,
the resulting likelihood that both \mgii\ systems correspond to
only one out of the six clusters is $\simeq 10^{4}$ times 
lower than the case where the individual absorptions are associated to
two different clusters (out of the available $6$).
This calculation corresponds to the average impact parameter $b_m=1200$h$_{71}^{-1}$kpc, characterising
this LOS,
and considers the measured median masses
of the RCS clusters in our sample (in the case where the double hit and single hit
are equally likely, the former are only $\simeq 5$ times less likely to happen than the latter for
this LOS).  Therefore, 
from this point on we consider this case as two single hits
(Table 1 is constructed
accordingly).  This would not be the case if the multiple absorption system was consistent with
the photometric redshift of a single RCS cluster.  This particular LOS offers
an excellent laboratory for follow-up spectroscopy of absorbers.

\begin{figure}
\begin{picture}(230,220)
\put(0,0){\psfig{file=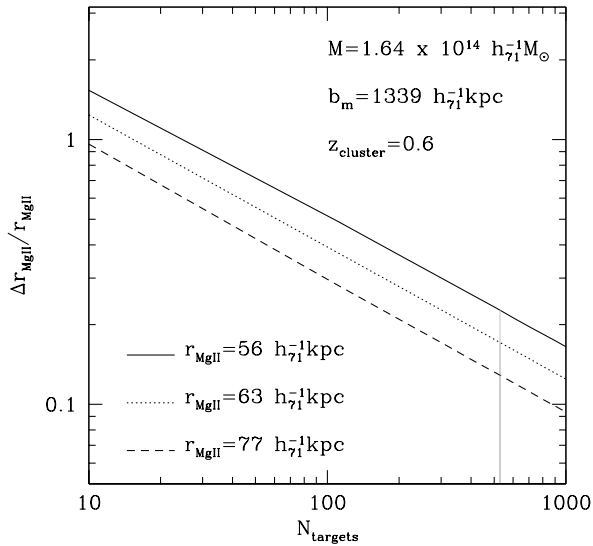,width=8.cm}}
\end{picture}
\caption{HO Model:
Relative errors in $r_{\rm \mgmath}$ as a function of number of target QSOs for
a fixed cluster mass of $M=1.64\times 10^{14}$h$_{71}^{-1}M_{\odot}$ and
median impact parameter $b_m=1339$h$_{71}^{-1}$kpc.  Results are shown for $r_{\rm \mgmath}=56$, $63$ and
$77$h$_{71}^{-1}$kpc.  The grey vertical line shows the number of 
QSO-cluster pairs in the S:St sample.
}
\label{fig:errors}
\end{figure}

We estimate the relative errors in \mgii\ halo sizes, $\Delta r_{\mgmath}/r_{\mgmath}$, assuming
Poisson statistics and
using the frequency of single hits and samples of different total number
of QSO-cluster pairs.  
For this calculation we adopt fixed values of
cluster mass,
$M=1.64\times 10^{14}$h$_{71}^{-1}M_{\odot}$, and impact
parameter, $b=1339$h$_{71}^{-1}$kpc, corresponding to the median values in our full sample of QSO-RCS cluster
pairs.  In Figure 2 we show the variation of $\Delta
r_{\mgmath}/r_{\mgmath}$ for three different \mgii\ halo sizes as a function of the
number of QSO-cluster pairs.   The vertical grey line indicates the expected errors
in the \mgii\ halo size for the total number of $529$ pairs in sample S:St,
and indicates that for sizes larger than $\simeq 50$h$^{-1}_{71}$kpc,
the uncertainties would be below a $25$ percent of the inferred size.

We will use this model to infer the typical size of the \mgii\ halo and its
associated measurement error
in the observational data described above.  

In what follows, we introduce the details of the additional model which uses a cosmological numerical
simulation in combination with a semi-analytic model of galaxy formation.

\subsection{GALFORM Semi-analytic galaxies}

This model differs from the HO in that it follows the evolution of galaxies using a number
of physical prescriptions for the different galaxy components, including cold gas, stars,
metallicity of stars and gas (hot and cold), and the properties of the 
central super-massive black hole.  However,
there is a strong observational component to the model since its parameters are tuned to
match the observed galaxy population at various redshifts.

The main advantage in using a semi-analytic model relies on the following.
The global properties of large concentrations of matter either
in simulations or in observations
are well known on average; however, the complex nature
of the non-linear collapse that formed virialised objects has important consequences on their
diversity, even for objects of similar masses.  The
properties that are not included in a general description of clusters include
their asphericity, substructure and formation history, and their relation to the surrounding environment.
Therefore, the use of a cosmological numerical simulation allows us to include this
diversity in the population of cluster-size dark-matter haloes, an effect that
is very difficult to include in the model presented in the previous subsection,
which relies on Monte-Carlo modeling.

In this work we use the Bower et al. (2006) GALFORM galaxy catalogue
which conforms a complete sample of galaxies down to a magnitude $M_r=-17$.
These galaxies populate the 
Millennium Simulation (Springel et al., 2005), which follows
the evolution of $\simeq1\times10^{10}$ collisionless dark-matter particles from $z=50$ to the 
present on a box of $704.2$h$^{-1}_{71}$Mpc a side, for a mass resolution per particle
of $1.21\times 10^9$h$^{-1}_{71}$M$_{\odot}$.  This simulation allows to identify bound
dark-matter structures and to follow their descendants at different redshifts.  This,
in turn, can be used to infer the galaxy population within each dark-matter halo,
via various assumptions regarding the different processes that shape the formation
and evolution of galaxies in a semi-analytic model (Croton et al., 2007, Malbon et al., 2007,
Lagos, Cora \& Padilla, 2008, De Lucia et al., 2006). 

It is important to note that these galaxies do not include a description of their
\mgii\ content, and therefore can only be used in a similar way to the HO model procedure.
Given that the model follows the evolution of the disk scale-length, $r_{\rm disk}$ of each
individual galaxy with time, we assume that the radius of the \mgii\ halos will be,
\begin{equation}
r_{\rm \mgmath}=A\times r_{\rm disk},
\label{prop}
\end{equation}
where $A$ is a proportionality constant that will be varied in order to match the observed frequency of
\mgii\ absorption events.
This is an important hypothesis that we will adopt throughout this work, namely, that the
size of the \mgii\ halo is proportional to the disk size.

Our sample of simulated clusters is constructed by selecting all systems above
a minimum mass such that the median mass equals that present in our RCS cluster sample.

\subsubsection{Simulated composite cluster}
\label{comp}

In order to study the frequency of galaxies in front of background QSOs in the
numerical model, we construct a composite cluster using all the objects within our
sample of simulated clusters.  We do this by stacking all the clusters using their most bound particle as their
centres; the semi-analytic model by Bower et al. (2006) uses this position to place
the central galaxy of each halo.  The stacking procedure erases all cluster membership
information.  Therefore our simulated composite cluster cannot be used to study 
multiple absorption events from individual clusters.  As was mentioned above, the
multiple hit candidate present in the data from paper I is considered as two single hits due
to its low probability of occurrence.

Given that the population of background QSOs is at a very large distance behind the
clusters in our sample, we assume that the former are distributed at random within $2$h$_{71}^{-1}$Mpc
from the cluster centres, in projection.  

\subsubsection{Morphological types}
\label{ssec:morph}

Among the information available for each galaxy in the semi-analytic model
are the stellar mass in the disk and bulge components.  Therefore, it is possible
to separate a population of elliptical and spiral galaxies, which may prove important
in light of previous results suggesting the presence of strong 
\mgii\ absorption systems mainly in spiral systems
(Zibetti et al., 2007).

Therefore, our analysis of number of galaxies in the line-of-sight of background QSOs
is done to a first approximation only on 
spiral galaxies selected so that the stellar content in their bulges is
only up to $70$ percent of the total stellar mass (Bertone et al., 2007)\footnote{
Bertone et al. find that in the case of the semi-analytic model we are using in this work, 
a limit of $70$ percent of the
total mass in stars is the minimum fraction for early type objects that allows to
reproduce the observed dependence of morphological fractions as a function
of stellar mass and local density.
}. 

We bear in mind that Zibetti et al. (2007) also show that there may
be dependencies of the \mgii\ absorption on galaxy colour and morphology.  Given the simplicity
of the approach we follow in this work, we will not attempt to model this dependency
but instead will offer estimates on the expected impact 
of the morphological selection on our results in Section \ref{ssec:sys}. 

\section{Results}

In this section we provide separately the results obtained using the two models
presented in the previous section, firstly from the observationally motivated Halo Occupation
model, and secondly from the semi-analytic model galaxies in the Millennium simulation (Springel et al., 2005).

\begin{figure}
\begin{picture}(230,220)
\put(0,0){\psfig{file=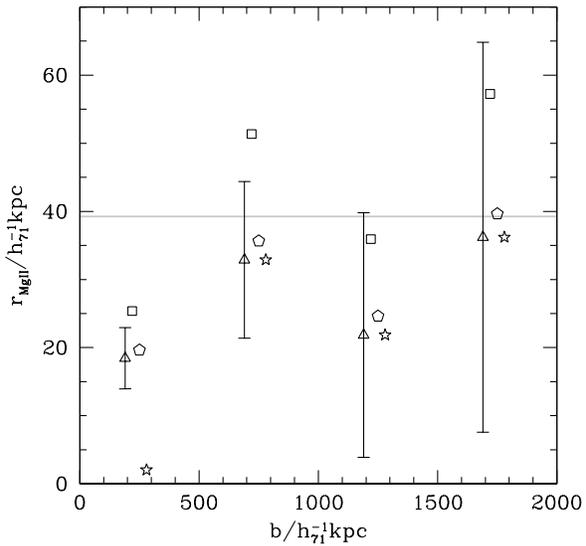,width=8.cm}}
\end{picture}
\caption{HO Model:
\mgii\ halo sizes as a function of impact parameter, according to the HO
model results.
Open squares show the results for the total number of hits in the S:WS sample,
open pentagons correspond to the number of hits corrected by the expected field
hits, and open triangles show the results from adding the correction for clustering.
The open star symbols indicate the resulting sizes when 
one of the cluster members is positioned at the centre of mass of the cluster.
The horizontal solid line indicates the radius of \mgii\ haloes in the field obtained
from the HO model.
}
\label{fig:gxsizes}
\end{figure}

\begin{figure}
\begin{picture}(230,220)
\put(0,0){\psfig{file=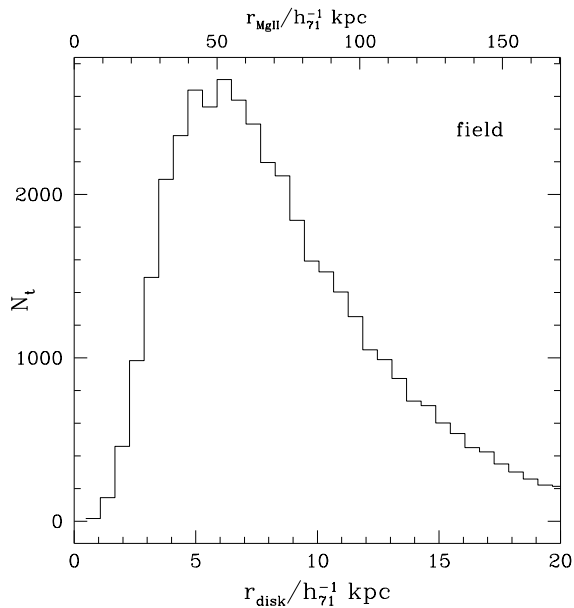,width=8.cm}}
\end{picture}
\caption{
Semi-analytic model:
Histogram of galaxy disk scale-lengths (lower x-axis) and \mgii\ galaxy haloes (upper x-axis) needed to reproduce
the number of hits reported for the field. 
}
\label{fig:fieldsizes}
\end{figure}

\subsection{HO Model}

We use Eq.  \ref{rmgii} to determine the sizes of
galaxy \mgii\ haloes that would produce the rate of hits shown in Table 1 (only calculated for the S:WS 
sample),
for clusters with median mass 
$1.64\times 10^{14}$h$_{71}^{-1}M_{\odot}$.
We show the resulting typical sizes in Figure \ref{fig:gxsizes} for the total number of hits
in the S:WS sample (open squares) and also for the corrected counts by the expected
contamination from the field (open pentagons) and clustering (open triangles).  
The results when considering the "farthest cluster'' variant of our S:WS sample
fall within the errorbars of these estimates, and therefore are not shown.
Errorbars are calculated by assuming Poisson statistics.
The points along the abscissa corresponds to the middle of the bin in $b$, with small shifts for the different
cases, added to improve clarity.

An important change in the inferred \mgii\ halo sizes is produced when one
of the cluster members is placed at the centre of the cluster
(i.e. a central galaxy).  This affects 
the innermost region for which it is straight-forward to infer the expected
fraction of line-of-sights that would pass through the central galaxy \mgii\ halo once its
typical size, $r_{\mgmath}$, is known.  This is an iterative process which converges rapidly after
three iterations.
When taking this into account, the results change quite dramatically for the low impact
parameter bin as it is shown by the open star symbols in Figure \ref{fig:gxsizes}.  These results indicate 
that near the cluster centres, the \mgii\ halo of a galaxy
could either be as large as $r_{\mgmath}\simeq 16$h$_{71}^{-1}$kpc, 
or as small as $r_{\mgmath}\simeq 2$h$_{71}^{-1}$kpc, depending on whether a central
galaxy is assumed to lie exactly at the observationally determined cluster centre (which
is subject to uncertainties). 

Our results show that the typical size of the \mgii\ halo tends to increase towards the outskirts of the cluster.  This
result is independent of the corrections applied to the observational data, and is not sensitive to
the presence of a central galaxy at the centre of mass of the cluster.
This is also in reasonable agreement with previous studies that report that the cold/cold-warm gas haloes
around galaxies may be stripped off by the hot intracluster gas (see for instance Heinz et al., 2003).

\begin{figure}
\begin{picture}(230,220)
\put(0,0){\psfig{file=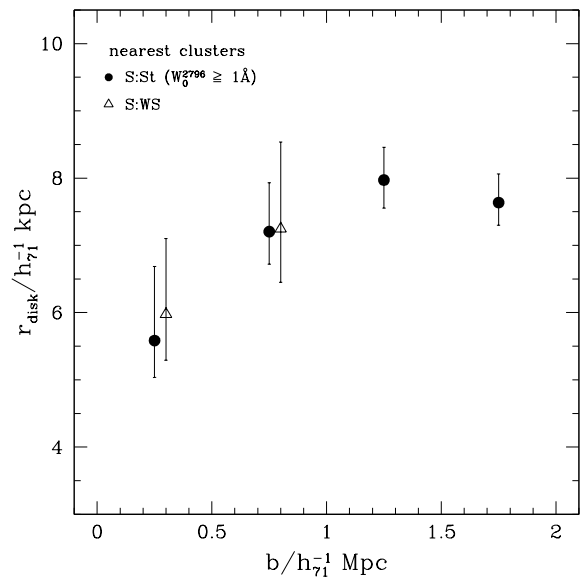,width=8.cm}}
\end{picture}
\caption{
Semi-analytic model:
Dependence on impact parameter to the cluster centre of the median disk scale-length in the
semi-analytic model.  The errorbars show the error of the median.
}
\label{fig:medrdisk}
\end{figure}

\begin{figure*}
\begin{picture}(430,360)
\put(30,180){\psfig{file=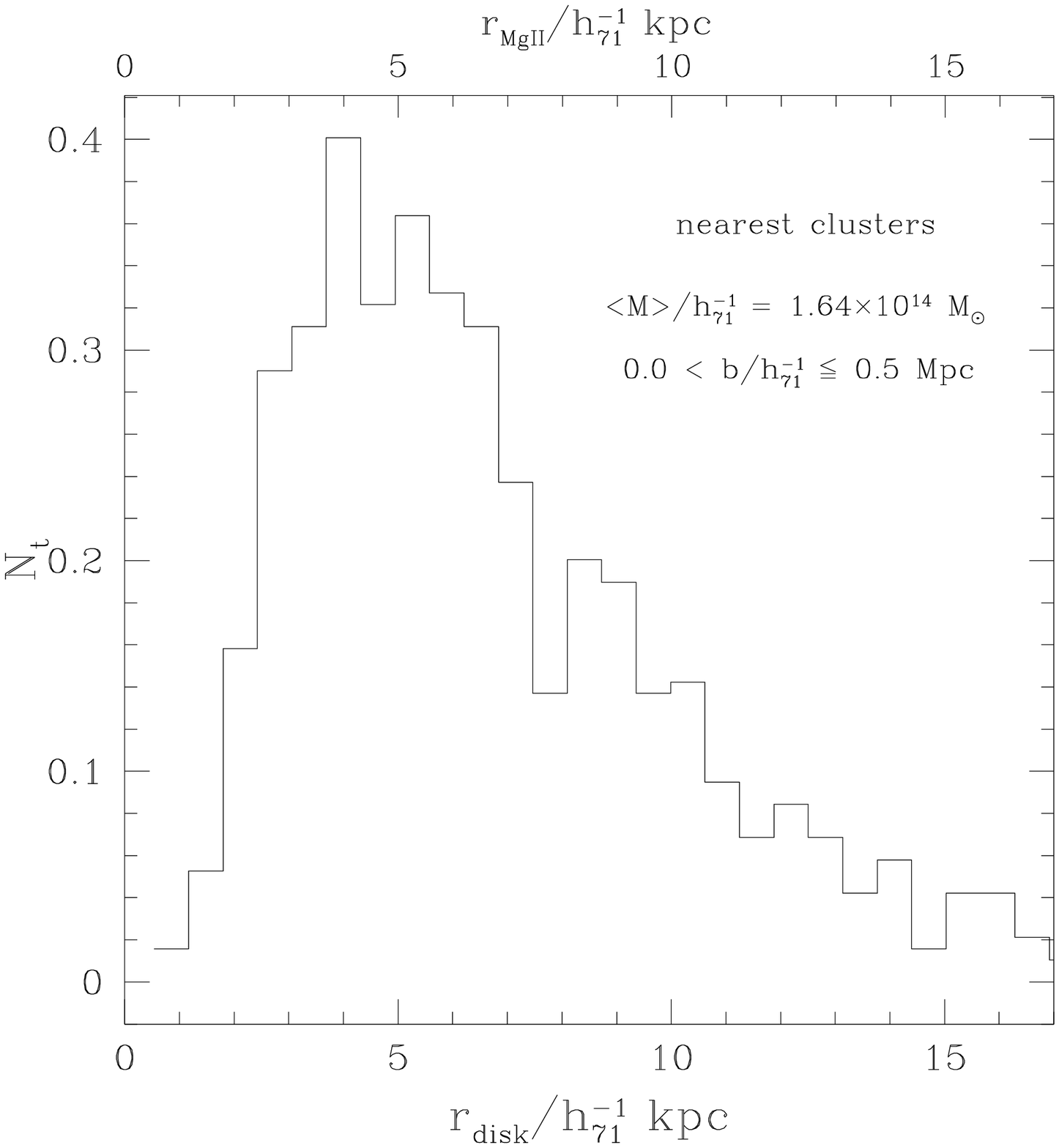,width=6.cm}}
\put(210,180){\psfig{file=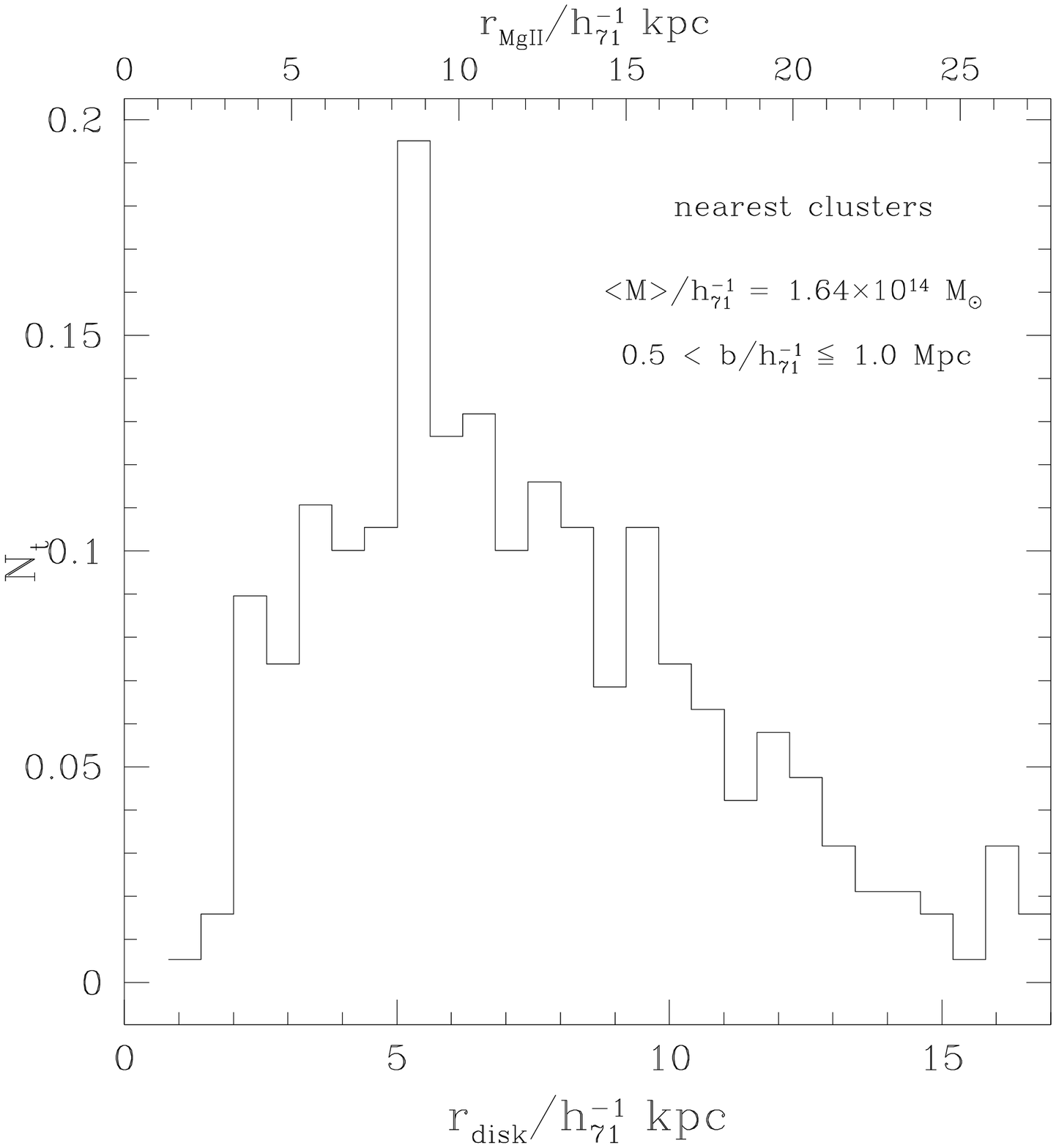,width=6.cm}}
\put(30,0){\psfig{file=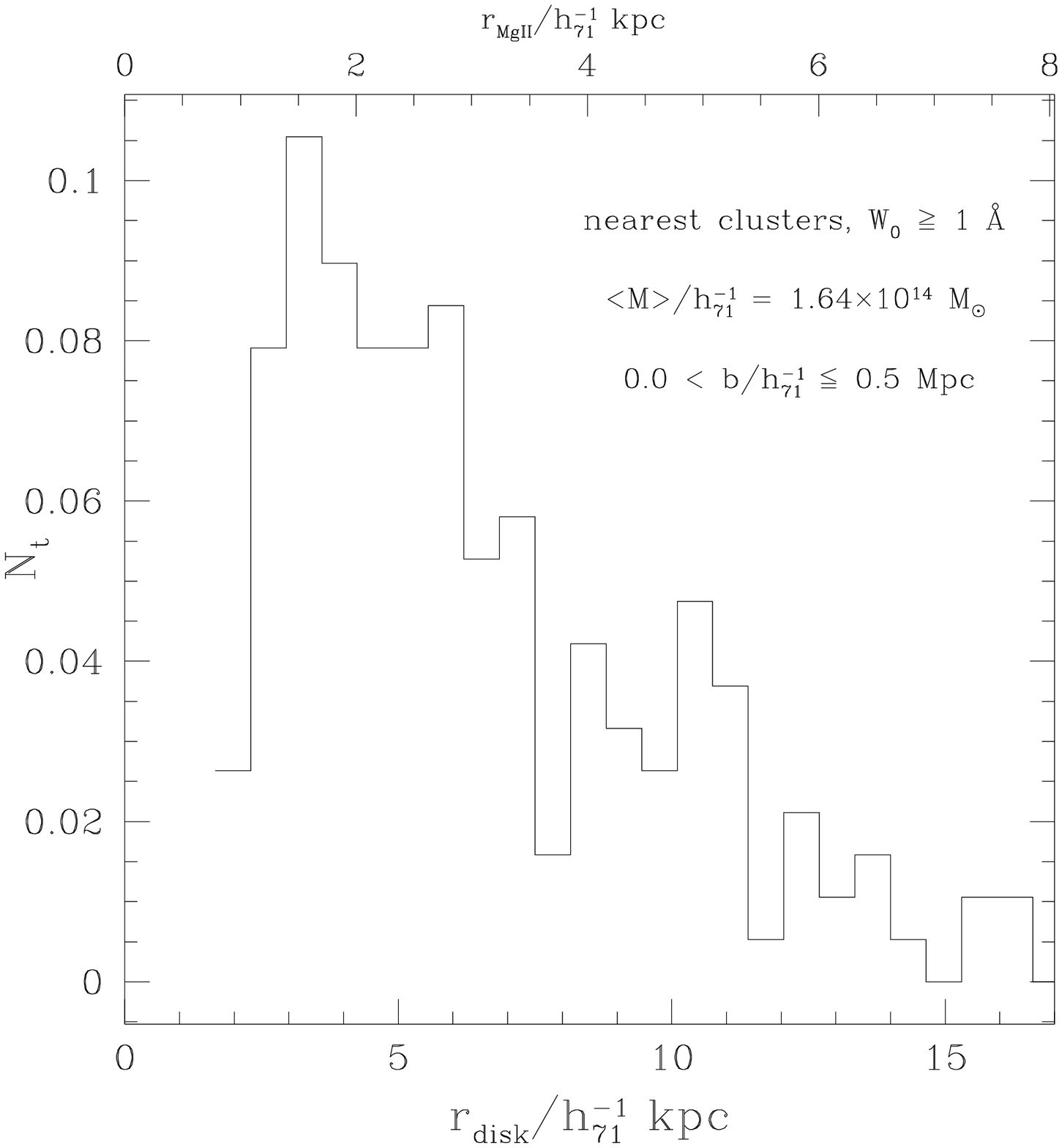,width=6.cm}}
\put(210,0){\psfig{file=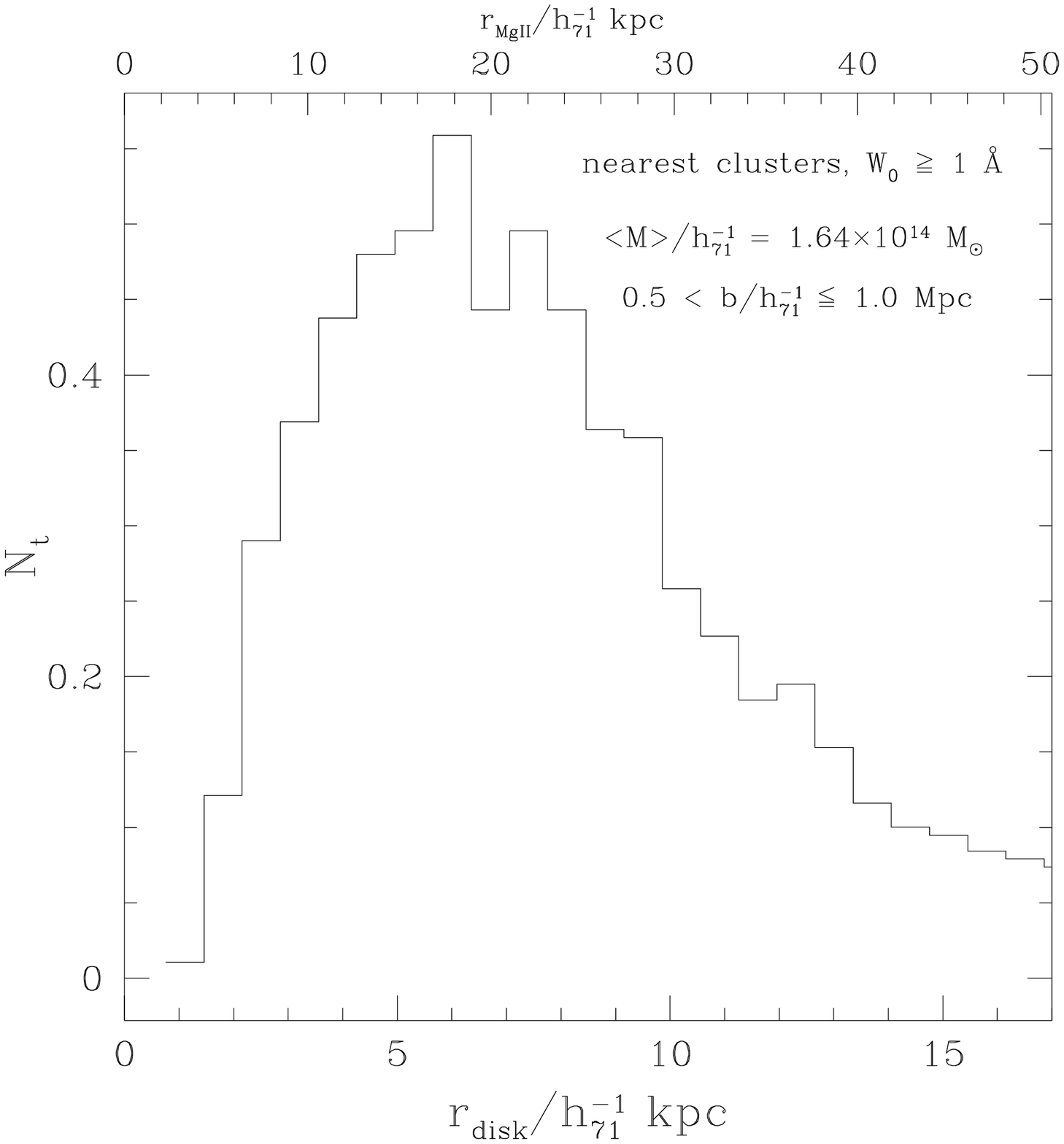,width=6.cm}}
\end{picture}
\caption{
Semi-analytic model:
Histogram of galaxy disk scale-lengths (lower x-axes) and \mgii\ galaxy haloes (upper x-axes) needed to reproduce
the number of hits in the observations, for different ranges of
impact parameter $b$, shown in the key (increasing values from left to right panels) for the S:WS and
S:St samples (top and bottom panels, respectively).
}
\label{fig:moresizes}
\end{figure*}

We also use the HO model to estimate the typical \mgii\ halo size in the field, which is
indicated by the horizontal grey solid line in Figure \ref{fig:gxsizes}. 
We estimate this size by integrating over the full range of halo masses where dark-matter
haloes are expected to have at least one galaxy with $M_r<-17$, for the EW range corresponding
to sample S:WS.
As can be seen, our estimate of $r_{\rm \mgmath}\simeq39$h$_{71}^{-1}$kpc (in broad agreement
with observational estimates, e.g. Churchill et al., 1999)
is slightly larger than our findings for the inner cluster environment,
in agreement with the stripping scenario in clusters.
For reference, the galaxy number density in the HO model is $n=7.639\times10^{-3}/$h$_{71}^{-3}$Mpc$^3$.
Notice that this
result corresponds to the HO model; the results from the analysis of the semi-analytic model for the
field are presented in the following subsection.

\subsection{GALFORM semi-analytic model}

Given the better description of individual clusters allowed by the combination
of a numerical simulation and a semi-analytic model, the results from the composite
cluster may provide further clues regarding the dependence of \mgii\ halo on the distance
to the cluster centre.

We start by studying the expected typical halo size in the field.  In order to do this, we randomly
select positions in the simulation and take all the galaxies within $2$h$_{71}^{-1}$Mpc
in projection from these centres stacked together.  This stack is characterised by
a comoving length which we convert to a redshift path.
We then place one QSO in the background and sort all the galaxies
in the stack with respect to their projected distance to the QSO, and assume that their \mgii\
halo sizes will be directly proportional to their disk scale-lengths, as was mentioned in Section 3.3.
All the \mgii\ haloes defined this way that contain the LOS to the QSO are counted.
We then vary the constant $A$ until the simulation reproduces the same counts per unit redshift path,
$dN/dz$, reported by the study of the field by Churchill et al. (1999)

Figure \ref{fig:fieldsizes} shows the results of the semi-analytic approach.  It
depicts the distribution of disk scale-lengths
in the field (lower x-axis label), where galaxies show a wide range of values
with a peak at $r_{\rm disk}=6$h$_{71}^{-1}$kpc.  The top x-axis label shows
the resulting \mgii\ halo sizes, with a distribution peak at a value of
$r_{\rm \mgmath}\simeq 50$h$_{71}^{-1}$kpc, well in agreement with the observational estimate
of Churchill et al. (1999) for EW$>0.02$\AA.
Also note that our estimate from this approach is only slightly higher than 
the results from the HO presented in the previous section.  This difference
arises from the different
EW limits assumed in each case (otherwise both results agree with
each other).

The analysis of the composite cluster constructed from the GALFORM model 
also provides useful insights on the size of the \mgii\ haloes in the cluster
environment, and particularly as a function of distance to the cluster centre.
In the semi-analytic model, galaxy disk scale-lengths show smaller
typical sizes near the cluster centres as can be seen in Figure \ref{fig:medrdisk}, where we
show the median size of galaxy
disks (points; errorbars indicate the error of the mean assuming Poisson statistics).
This is a first indication that
the model indeed includes important environmental effects on the galaxy population (notice
that these values are not to be compared to the field \mgii\ region sizes denoted as $r_{\rm \mgmath}$).
In the GALFORM semi-analytic model, the physical process driving the environmental changes within clusters is
the removal of the hot gas reservoir around galaxies once they enter a new dark-matter halo; this process
effectively stops the cooling of gas onto the galactic disk halting its growth.

\begin{figure}
\begin{picture}(230,220)
\put(0,0){\psfig{file=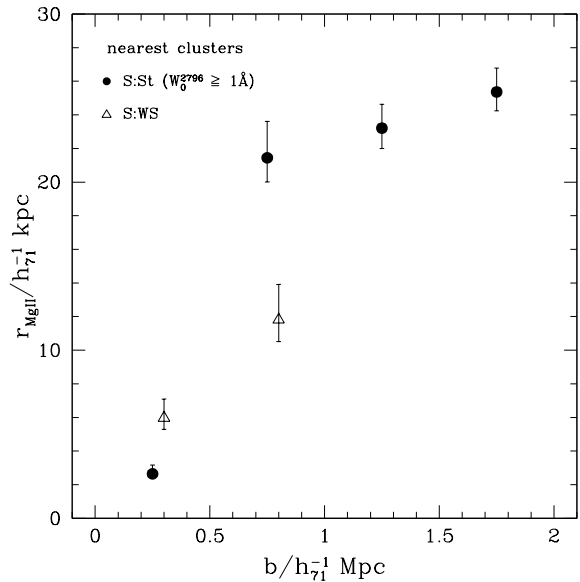,width=8.cm}}
\end{picture}
\caption{
Semi-analytic model:
Median sizes of \mgii\ haloes as a function of projected distance to the cluster centre.  Errorbars
show the error of the median.
}
\label{fig:mgii}
\end{figure}

Following
the procedure used to infer the typical sizes of \mgii\ haloes in the field, we also
find the proportionality constant that multiplied by the disk scale-lengths
gives a \mgii\ cloud size that provides a match to the abundance
of \mgii\ absorption systems in cluster-QSO pairs in the S:WS and S:St samples.  
Figure \ref{fig:moresizes} shows histograms of galaxy disk scale-lengths (lower x-axes)
and resulting typical \mgii\ halo sizes (top x-axes)
for two ranges of impact parameter (increasing values of $b$ from the left to the right panels)
for the S:WS (top) and S:St (bottom) samples.
The size distributions show a marked raise towards a single peak 
(on the position of these peaks we will centre
our discussion) and a slower fall-off for higher values of $r_{disk}$,
with a distribution width that clearly increases towards the cluster outskirts, in
both disk and \mgii\ halo sizes.  From the location of the peaks it can be seen that
the increase of the disk scale-lengths
with the impact parameter shown in Figure 5 is confirmed.  It can also be seen that 
the location of the
\mgii\ size distribution peaks show a similar trend of larger values further away from
the cluster centres.  Furthermore,
the \mgii\ halo sizes are consistent with larger values for the S:WS sample for
the lowest impact parameter bin,
which reflects the fact that the frequency of \mgii\ absorption events is higher in this sample
than in S:St at such distances from the cluster centre.

Figure \ref{fig:mgii} shows the median size of a \mgii\
cloud as a function of the distance to the cluster centre (errorbars show the
error of the median) for the S:WS (triangles) and S:St (circles) samples.
As can be seen from both figures, the trend of a larger galaxy size
at larger distances from the cluster centre is clearly found in both samples
with values ranging from $r_{\rm \mgmath}\simeq 2-6$h$_{71}^{-1}$kpc at
$b<0.5$h$_{71}^{-1}$Mpc
to
$r_{\rm \mgmath}\simeq 25$h$_{71}^{-1}$kpc for
$b>1.5$h$_{71}^{-1}$Mpc, in good agreement with the HO results with a central galaxy
\footnote{Notice that the \mgii\ halo size at $b>1.5$h$_{71}^{-1}$Mpc 
is not expected to resemble the \mgii\ halo size inferred for the field 
since this sample contains only galaxies in high mass dark-matter
haloes, and therefore does not include a representative sample of field galaxies, even in the cluster
outskirts.}.
The corresponding inferred size for typical \mgii\ haloes in the field for the
range of EW in sample S:WS for the semi-analytic model is $40$h$_{71}^{-1}$kpc (notice that
this value differs slightly with the result from Figure 4, due to the different EWs considered).

We remind the reader
that these results are motivated by the observational frequency of \mgii\ absorption systems
in our samples of RCS cluster-QSO pairs (the proportionality constant
in Eq. 10 depends on the impact parameter).  Therefore, the trends of smaller \mgii\ halo (cf. Fig. \ref{fig:mgii}) and disk 
sizes (cf. Fig. \ref{fig:medrdisk}) towards the cluster centres are independent results.
Furthermore, the reasonable consistency between the results from the HO model and
the semi-analytic simulation provides further reliability to the findings on the trends
of \mgii\ halo size with the distance to the cluster centre.

\section{Discussion}

In the previous section we studied the inferred typical \mgii\ halo sizes using two different models, the HO and
the output from a semi-analytic simulation.  We find that both models are able to reproduce the observed frequency
of \mgii\ absorption events found in observational data in Paper I.  More importantly,
our results indicate that the \mgii\ halo sizes
tend to increase from relatively low values of
$r_{\rm \mgmath}\simeq 2-6$h$_{71}^{-1}$kpc to $r_{\rm \mgmath}\simeq 26-40$h$_{71}^{-1}$kpc 
for $b=0.5$h$_{71}^{-1}$Mpc to
$b>1.5$h$_{71}^{-1}$Mpc; both models produce consistent results.

In what follows, we present a study of the fraction of the \mgii\ halo responsible for producing different
\mgii\ absorption line strengths, and an analysis of possible systematic effects 
present in our current measurements.

\subsection{Regions of strong \mgii\ absorption line systems}

We take full advantage of the additional information available in the semi-analytic
model regarding the galaxy disk scale-lengths, and combine this with previous results by 
Chen \& Tinker (2008) who find that stronger absorption systems are produced
by clouds closer to the galaxy centres\footnote{In their terminology this would be a halo
centre, which can be applied to each galaxy in the semi-analytic model since these
are located at the centres of bound substructures (or sub-haloes) in the numerical
simulation.} to study the typical size of the regions in the \mgii\ halo
responsible for different absorption line strengths.

\begin{figure}
\begin{picture}(230,220)
\put(0,0){\psfig{file=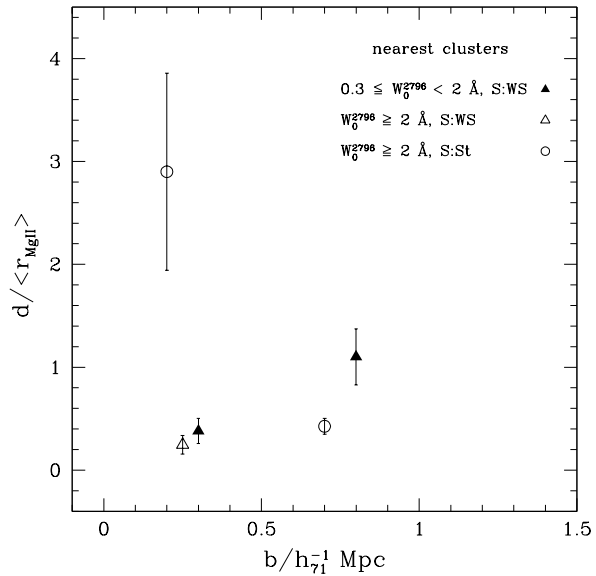,width=8.cm}}
\end{picture}
\caption{
Semi-analytic model:
Extent of different absorption line equivalent widths in units of the \mgii\ halo radius,
as a function of impact parameter to the cluster centres.
The results from the S:WS sample are shown by triangles, and from the S:St sample by circles.
Open symbols correspond to strong absorption line systems; filled symbols to weak absorption lines.
}
\label{fig:gxsizes2}
\end{figure}

In our analysis of the semi-analytic model we maintain a record
of the distance, $d$, between each QSO line-of-sight and each galaxy centre.  Therefore, after
finding the proportionality constant $A$ (Eq. \ref{prop}) that reproduces the frequency of
hits found in Paper I,
we can sort the galaxies with respect to $d$. 
Then, for different ranges of cluster-centric impact parameter, we study the observed fractions of different
line-strengths, and find the galacto-centric distance out to which the same fractions of model
galaxies are included.  Such a distance is then identified as the maximum distance out to which
a given line-strength would be produced in the model.
Figure \ref{fig:gxsizes2} presents
the resulting fraction of the \mgii\ halo size out to which a given line-strength occurs,
as a function of distance to the cluster centre.  This rescaling is adopted in order
to factor out the measured variation of the global \mgii\ halo size with the 
impact parameter to the cluster centre.
Open and solid symbols represent high ($W_0^{2796}>2$\AA) and low ($W_0^{2796}<2$\AA) equivalent width absorption systems, respectively;
triangles correspond to results from the S:WS sample, circles to S:St.
In the
case of the high equivalent width results, the large errorbars only allow a possible rejection of an increase
of the \mgii\ halo region producing such absorption events.
However, there is a mild indication that the low equivalent width absorption systems (all from S:WS) tend
to occur at increasingly larger distances from the galaxy centre in terms of the \mgii\ halo size towards
the outskirts of clusters.  These
results are again
in good agreement with the stripping scenario, where the low density \mgii\ clouds within a given galaxy
would be more easily removed
than the denser \mgii\ regions nearer the galaxy centre
that would be responsible for high equivalent width absorption systems.

\subsection{Possible systematic effects}
\label{ssec:sys}

The parametrisation of both models, HO and semi-analytic, used to mimic the
observational procedure carried out in Paper I, depends on a number of parameters,
assumptions and measurements that can induce systematic effects.

For instance, the results shown in the previous sections were obtained for clusters of a median
mass corresponding to the estimates for the RCS clusters.  
In order to assess the possible systematic biases arising from errors in these cluster mass measurements
we change the adopted median mass by factors of $4$ and $0.25$,
approximately $0.6$dex in the cluster mass.  The results in the HO and semi-analytic
models are affected by very similar relative variations, which do not show a dependence with
the distance to the cluster centre.  When the lower value for the median mass is adopted, the inferred
median sizes of the \mgii\ haloes increases by $\sim 10$ percent; alternatively, a higher median mass
produces a decrease in the inferred size by a $20-25$ percent.  Given that the effect does not vary
with the distance to the cluster centre, our conclusions would not be significantly affected
by this source of systematic uncertainties.

It should also be taken into account that in our modeling, we have considered
spherical \mgii\ haloes.  Given the lack of detailed information on the
thickness of cold gas disks, we do not venture into assuming a given disk thickness in this
work, as has also not been done in previous works studying the size of \mgii\ absorption regions
in the field as in Churchill et al. (2001).


It has been argued that a unity covering fraction for \mgii\ haloes is rather unlikely
(Bechtold \& Ellingson, 1992, Tripp  \& Bowen, 2005, Chen \& Tinker, 2008, Kacprzak et al., 2008).
Since the models we have adopted here assume a fixed value for the covering fraction 
(which we have chosen to be $C=1$), we cannot  study a possible dependence with 
environment. However, it is easy to see that a different covering fraction 
will change our results on the MgII halo sizes according to
$r_{\mgmath}$ to $r_{\mgmath}^C=\frac{1}{\sqrt{C}}r_{\mgmath}$.

Finally, the effect from considering absorption from low 
gas content galaxies should also be taken into account.
In the semi-analytic model, the fraction of elliptical galaxies changes as a function of the distance
to the cluster centre.  This is specially true 
for $r/r_{vir}<1$, where the fraction of model elliptical
galaxies (see Section \ref{ssec:morph}) increases from $\sim 15$ percent to $\sim 50$ percent at
$r/r_{vir}=0.015$, measured in projection (these fractions are in good agreement with results from
the SDSS from Goto et al., 2003).  At larger distances from the cluster centres, 
the fraction of elliptical
galaxies remains practically unchanged.  For sample S:WS
(and similarly for sample S:St), considering an absorption from elliptical
galaxies in the semi-analytic model would increase by a factor of $\sim 1.43$(\footnote{
This is an upper limit for the contribution of ellipticals to the projected galaxy number density
at a median impact parameter of $b/r_{vir}=0.17$.  This value corresponds to the first bin in our measurements of $r_{\mgmath}$,
$b<500$h$_{71}^{-1}$kpc, for the median virial radius of the clusters, $r_{vir}=775$h$_{71}^{-1}$kpc.})
the number of potential
absorbers and therefore decrease our estimate of the typical \mgii\ size inside clusters by a $16.4$ percent.  Our
estimates at larger distances from the cluster centre will be equally affected by a decrease of
approximately $7.8$ percent.  The latter estimate would also change our estimates
for the typical \mgii\ halo size inferred for the field accordingly in both, the semi-analytic and HO models.  
In consequence, our conclusions do not change significantly due to this effect.

\section{Summary and conclusions}

In this paper we have presented a theoretical study of \mgii\ halo sizes of galaxies in clusters
and in the field, by comparing recent observational results on the incidence of
\mgii\ absorption systems in RCS cluster-QSO pairs by Lopez et al. (2008), and the predicted
systems from the Halo Occupation and Semi-analytic models.

As a first step in our analysis we applied corrections to the observed frequency of
absorption events to take into account possible contaminations from the field, and effects
of the Large-Scale Structure.

The results from the HO and semi-analytic models indicate that the typical 
field \mgii\ halo is
$r_{gal}\simeq 39-50$h$_{71}^{-1}$kpc (for EW$>0.02$\AA), a value in broad agreement
with previous results pointing towards
$r_{gal}\simeq 50-100$h$_{71}^{-1}$kpc found for similar absorber EWs
(Churchill et al., 2001, Zibetti et al., 2007, Chen \& Tinker, 2008, Kacprzak et al., 2008).  

The HO model
also indicates that the \mgii\ halo tends to be smaller near the cluster centres, reaching values of
$r_{gal}\simeq 16$h$_{71}^{-1}$kpc.  When one of the cluster members occupies
the centre of mass of the cluster, the \mgii\ halo size can be as low as
$\simeq 2$h$_{71}^{-1}$kpc.  
This result
is obtained assuming a fixed galaxy size for all the cluster members.
The results from the Semi-analytic model are consistent with those from the HO
showing a clear
decrement in the typical \mgii\ size towards the cluster centres.  
An interesting by-product of this analysis is that the galaxy disk scale-lengths in
the model also show a clear increase towards the outskirts of clusters, in accordance with
the scenario where strong interactions strip galaxies of their outer disks in cluster environments.

We also used the semi-analytic model to estimate the radii of \mgii\ haloes producing
different strengths of absorption.  In order to do this we assumed that stronger lines are produced nearer
the centres of galaxies (following Chen \& Tinker, 2008).  We found that in terms of the
\mgii\ halo size, the strong absorption regions occur out to a rather fixed value, independent
of the distance to the cluster centre.  However, weaker absorption line systems show a tendency to occur
at larger distances from the galaxy centre (in units of their global \mgii\ halo radius) towards
the outskirts of clusters.
This trend could explain the flatter EW distribution found in paper I, that included all impact 
parameters within $b<2$h$_{71}^{-1}$Mpc.

Our results indicate that the effect of the cluster environment on galaxies is extremely
important and can produce a decrement of up to a $90$ percent of the \mgii\ halo size with respect
to galaxies in more gentle environments such as the field.

\section*{Acknowledgments}
SL, LFB, PL and NP were partly supported 
by the Chilean Centro de Astrof\'\i sica FONDAP No. 15010003. NDP was also supported by
FONDECYT grant No 1071006, SL 
by FONDECYT grant No 1060823, and LFB by FONDECYT grant No 1085286.
We thank the anonymous referee for helpful comments and suggestions.
Funding for the SDSS and SDSS-II has been provided by the 
Alfred P. Sloan Foundation, the Participating Institutions, the National Science Foundation, 
the U.S. Department of Energy, the National Aeronautics and Space Administration, the 
Japanese Monbukagakusho, the Max Planck Society, and the Higher Education Funding 
Council for England. The SDSS Web Site is http://www.sdss.org/.

\bsp

\label{lastpage}

\end{document}